\documentclass[prb, reprint, superscriptaddress,floatfix]{revtex4-2}

\usepackage{amsmath}
\usepackage{graphicx}
\usepackage{verbatim}
\usepackage{xcolor}
\usepackage{hyperref}

\begin{document}

\title{Magnetic Photocurrents in Multifold Weyl Fermions}
\author{Sahal Kaushik}
\email{sahal.kaushik@stonybrook.edu}
\affiliation{Department of Physics and Astronomy, Stony Brook University, Stony Brook, NY 11794, USA}
\author{Jennifer Cano}
\email{jennifer.cano@stonybrook.edu}
\affiliation{Department of Physics and Astronomy, Stony Brook University, Stony Brook, NY 11794, USA}
\affiliation{Center for Computational Quantum Physics, Flatiron Institute, New York, NY 10010, USA}

\begin{abstract}
    We examine the magneto-optical response of chiral multifold fermions. Specifically, we show that they are ideal candidates for observing the Helical Magnetic Effect (HME) previously predicted for simple Weyl fermions. Unlike Weyl fermions, the HME is present in multifold fermions even in the simplest case where the low-energy dispersion is linear and spherically symmetric.
    In this ideal case, we derive an analytical expression for the HME and find it is proportional to the circular photogalvanic effect;  for realistic parameters and accounting for the geometry of the setup, the HME photocurrent could be roughly the same order of magnitude as the circular photogalvanic effect observed in multifold fermions.
    Additional non-linear and symmetry-breaking terms will ruin the quantization but not hurt the observation of the HME.
\end{abstract}
\maketitle

\section{Introduction}
\label{sec:intro}
Weyl fermions in condensed matter physics are gapless chiral quasi-particles with (pseudo)-spin coupled to their momentum \cite{Wan11,Weng15,Huang15,Xu15,Lv15,Xu15a,Lv15a}.
The chirality of fermions leads to interesting physical consequences such as surface Fermi arcs \cite{Wan11}, the Chiral Anomaly \cite{AnomalyAdler,AnomalyBJ,Xiong2015} and the Chiral Magnetic Effect (CME) \cite{CME}, which are accompanied by negative magnetoresistance \cite{son2013chiral,burkov2014chiral,kim2013dirac,Liang2015,Huang2015,Zhang2016,Wang2016,Arnold2016,li2016chiral,zhang2017room}.

The chirality of Weyl fermions also plays a role in their optical response.
For example, a helicity-dependent photocurrent due to partial Pauli blockade of tilted Weyl cones has been predicted \cite{PLee} and observed in TaAs \cite{MaTaAs}. Similar helicity-dependent photocurrents have also been observed in response to ultrafast near-infrared and optical pulses in TaAs \cite{SiricaTaAs,SahalTHz}. 

In asymmetric Weyl materials, additional types of optical response are possible.
Asymmetric Weyl materials lack a symmetry that relates Weyl cones of opposite chiralities. 
Thus, left and right handed fermions can have different energies and velocities and, consequently, interact differently with electromagnetic fields. 
For example, asymmetric Weyl materials are predicted to exhibit a quantized Circular Photogalvanic Effect (CPGE), i.e. a photocurrent in the direction of circularly polarized light, when Weyl cones of one chirality are fully Pauli blockaded \cite{deJuanQCPE}. The Pauli blockade is only possible when the left and right handed cones are at different energies, which is why this effect is specific to asymmetric Weyl materials. 

In this manuscript, we will study the Helical Magnetic Effect (HME), which predicts a photocurrent in the presence of a magnetic field in a tilted, asymmetric Weyl material \cite{YutaHME}. 
This effect can only occur in the absence of inversion and any mirror reflection (which is possible only in asymmetric Weyl materials) and in the absence of the product of inversion and particle-hole symmetry (which is possible only if the Weyl cones are tilted). It is not enough to have a finite chemical potential on untilted cones; the dispersion relation itself must lack the product of inversion and particle-hole symmetry.
A transverse magnetic photocurrent has also been predicted in symmetric Weyl materials with tilted cones such as TaAs \cite{GolubCircular, SahalTHME}.

A challenge in observing the quantized CPGE or the HME is the lack of asymmetric Weyl materials, which must also have tilted Weyl cones to exhibit the HME.
Asymmetric Weyl materials necessarily have a chiral crystal structure \cite{HasanSrSi2,HasanRhSi} or magnetic ordering \cite{ray2020tunable} (although it is also possible to engineer an asymmetric Weyl material by applying an external field \cite{SahalAsymm})
and these materials are relatively rare.
However, recently, asymmetric chiral multifold fermions have been discovered in certain compounds with the B20 crystal structure \cite{tang2017multiple,HasanRhSi,sanchez2019topological,schroter2019chiral,RhSiCurrent,xu2020optical}.
Multifold fermions are generalizations of Weyl and Dirac fermions that exhibit either a higher degeneracy or a different topology \cite{CanoMultifold,cano2019multifold}.
The chiral multifold fermions are also asymmetric and thus exhibit a quantized CPGE, which has been observed in experiment \cite{FlickerMultifold,RhSiCurrent,CoSiCurrent}. They have also been predicted to cause a quantized circular dichroism \cite{MandalDichroism}.
But since the known multifold materials occur at high-symmetry momenta where a tilt is forbidden by crystal symmetry, naively they should not exhibit the HME.

The purpose of this manuscript is to show that this naive expectation is incorrect: in fact, chiral multifold fermions are an ideal platform to exhibit the HME.
We show that chiral multifold fermions exhibit the HME even in the idealized limit where they have perfect spherical symmetry and a linear dispersion, as long as the Fermi level is not exactly at the degeneracy point. 
In this limit, the HME in multifold fermions takes a particularly simple form and is related to the quantized CPGE by a factor of the inverse of the number of Landau levels involved in the photoexcited transitions.
We plot the HME for both spin-1 and spin-3/2 fermions to explicitly demonstrate our results. 
Away from the idealized symmetric and linear limit, the HME is present, but non-linear and symmetry-breaking terms ruin its quantization.
We illustrate this for a double spin-1/2 fermion which splits into a spin-1 fermion and a trivial fermion by terms that break spherical symmetry; this example is relevant to the multifold fermions found in B20 compounds.
We end with a discussion of the relevance of these results to the experimentally characterized compounds CoSi and RhSi.


\section{Hamiltonians for Chiral Fermions}\label{secHam}

In this section, we review the Hamiltonian and Berry curvature of chiral symmetric and tilted Weyl and multifold fermions.
In the simplest incarnation, a Weyl cone is described by the continuum Hamiltonian
\begin{equation}
	H = \chi v_0\vec{k}\cdot\vec{\sigma},
	\label{eq:H0}
\end{equation}
where the chirality $\chi$ is $+1$ for a right handed cone and $-1$ for a left handed cone. This Hamiltonian yields a linear dispersion $E = \pm v_0k$ and a velocity $\vec{v} = \pm v_0\hat{k}$, where $\pm$ corresponds to the upper/lower band. The chirality of the Weyl cone in Eq.~(\ref{eq:H0}) can be also be defined as $\chi = \mathrm{sgn}(\vec{v}\cdot\vec{s})$ where $\vec{v}$ is the velocity and $\vec{s}$ the (pseudo)-spin of the fermion. This definition is valid for both bands of the Weyl cone: the upper/lower bands have opposite chirality corresponding to velocity aligned/anti-aligned with spin. 
An individual Weyl cone has electrons of fixed chirality; thus, an individual Weyl cone lacks inversion symmetry ($P$), which flips momentum but not spin.
Instead, in a crystal with inversion symmetry, the inversion symmetry operator will exchange Weyl cones of opposite chirality.

Eq.~(\ref{eq:H0}) has spherical symmetry, which is broken by the lattice.
More generally, a Weyl fermion can have tilt and anisotropy, and is described by the Hamiltonian:
\begin{equation}
	H = v^i_a k_i\sigma_a + v^i_t \sigma_0 k_i
	\label{tilted}
\end{equation}
where $v^i_a$ describes the untilted part of the Hamiltonian, which might be anisotropic, and $v^i_t$ describes the tilt. The chirality is $\chi = \mathrm{sgn}(\mathrm{det}\ v^i_a)$.

Weyl points are quantized monopole charges of Berry curvature. 
For a symmetric linear Weyl cone described by Eq.~(\ref{eq:H0}), the Berry curvature is of the form $\vec{\Omega} = \pm \chi \hat{k}/2k^2$, where $\pm$ correspond to the upper/lower band; more generally, the Berry curvature will be anisotropic. Whether isotropic or not, integrating the Berry curvature over a Fermi surface enclosing a single linear Weyl fermion of the form of Eq.~(\ref{eq:H0}) or (\ref{tilted}) yields $2\pi C$, where $C = \chi = \pm 1$ is the Chern number of the Fermi surface. By the Nielsen-Ninomiya theorem \cite{nielsen1983adler}, the total number of left and right handed cones in the Brillouin zone must be equal so that the total Berry flux vanishes.

Although Weyl fermions do not require any crystal symmetry, crystal symmetries can protect the following generalizations of Weyl fermions.
Rotation symmetries protect Weyl fermions with $|C|>1$, which have quadratic- or cubic-dispersions along certain directions \cite{fang2012multi,HasanSrSi2,FourWeyl}.

Chiral multifold fermions, which are higher-spin generalizations of Weyl fermions \cite{CanoMultifold,cano2019multifold,FlickerMultifold}, can be protected by symmetry in chiral nonsymmorphic crystals.
A spin-$J$ chiral multifold fermion has $2J+1$ bands. The simplest (spherically symmetric) Hamiltonian for such a fermion is
\begin{equation}
H = \chi v_0 k_i S^i
\label{eq:HMF}
\end{equation}
where $S_i$ are the spin-$J$ matrices. 
Spin-1 (three-fold degeneracy) and spin-3/2 (four-fold degeneracy) flavors are possible in 3D crystals.
In addition, double spin-1/2 (four-fold degeneracy) and double spin-1 (six-fold degeneracy) fermions can also be symmetry-protected.
The Hamiltonian of a double spin-$J$ fermion is of the form
\begin{equation}
H = \chi \tau_0 v_0 k_i S^i
\label{eq:DHMF}
\end{equation}
where the Kronecker product is implied and $\tau_0$ is the $2\times 2$ identity matrix acting on some additional degree of freedom outside of the spin-$J$ multiplet; at least one crystal symmetry must be off-diagonal in the basis of $\tau$ matrices for the multifold fermion to be symmetry-protected.

Since the Hamiltonians in Eqs.~(\ref{eq:HMF}) and (\ref{eq:DHMF}) are spherically symmetric,
each band can be labelled by the projection of spin along momentum, $S_k$. The Chern number of a Fermi surface in a band with spin-projection $S_k$ is $2S_k$; the integral of the Berry curvature over that Fermi surface is $2\pi\times 2S_k$.
As discussed for Weyl fermions, the total Chern number in the Brillouin zone must always be zero. Thus, materials with multifold fermions can have multifold fermions of both chiralities or a multifold fermion of one chirality and the appropriate number of simple Weyl fermions of the other chirality.

In the following, we focus on spin-1 and spin-3/2 fermions.
The calculation of the HME for double spin-$J$ fermions is the same as single spin-$J$ fermions with an additional factor of two.

\section{Symmetry and Magnetic Photocurrent}\label{secSym}

As mentioned in Sec.~\ref{sec:intro}, the HME requires a tilted Weyl cone.
To describe the tilt, we need to introduce charge-conjugation symmetry ($C$), which
maps one electron to another electron with opposite energy, momentum, and angular momentum, but same velocity. Since removal of an electron with velocity $\vec{v}$ and angular momentum $\vec{s}$ is equivalent to the creation of a hole with velocity $\vec{v}$ and angular momentum $-\vec{s}$,  charge-conjugation maps an electron to a hole with opposite chirality. 
This is analogous to the situation in high energy physics, where the antiparticle of a left-handed neutrino is a right-handed antineutrino. 

A \textit{single} untilted Weyl cone cannot have inversion symmetry (as discussed below Eq.~(\ref{eq:H0})) or charge-conjugation symmetry because these symmetries both flip chirality.
But a Weyl fermion whose low-energy bands are linear and whose dispersion relations of electrons and holes are similar, has \textit{approximate} $CP$ symmetry, which is broken by quadratic terms. 
This is analogous to the situation in the Standard Model, where the terms that break $P$ and $C$ are large, but the $CP$ violating terms are very small.

To reiterate, the energy, momentum, angular momentum, and velocity of \textit{electrons} transform under $P$, $C$, and $CP$ symmetries as follows:
\begin{align}
    P:\quad & E \to +E,\quad \vec{k} \to -\vec{k},\quad \vec{s} \to +\vec{s},\quad \vec{v} \to -\vec{v}\nonumber\\
    C:\quad & E \to -E,\quad \vec{k} \to -\vec{k},\quad \vec{s} \to -\vec{s},\quad \vec{v} \to +\vec{v}\\
    CP:\quad & E \to -E,\quad \vec{k} \to +\vec{k},\quad \vec{s} \to -\vec{s},\quad \vec{v} \to -\vec{v}\nonumber
\end{align}
The Hamiltonians of untilted simple Weyl cones (Eq.~(\ref{eq:H0})) and multifold fermions (Eqs.~(\ref{eq:HMF}) and (\ref{eq:DHMF})) satisfy $CP$, while the Hamiltonian of a tilted Weyl cone (Eq.~(\ref{tilted})) does not.

In the quantized CPGE \cite{deJuanQCPE}, circularly polarized light, characterized by its angular momentum $\vec{J}$, produces a current $\vec{j}$, related by:
\begin{equation}
\vec{j} = \beta_l \vec{J}
\end{equation}
Since the current is odd under both $P$ and $C$, and the angular momentum is even under both, the coefficient $\beta_l$ is also odd under $P$ and $C$ but even under $CP$. This is also how a single Weyl cone transforms under $P$ and $C$. This is why the quantized CPGE occurs when there is an imbalance of occupied Weyl cones of each chirality.

\begin{figure}
\centering
\includegraphics[scale=0.24]{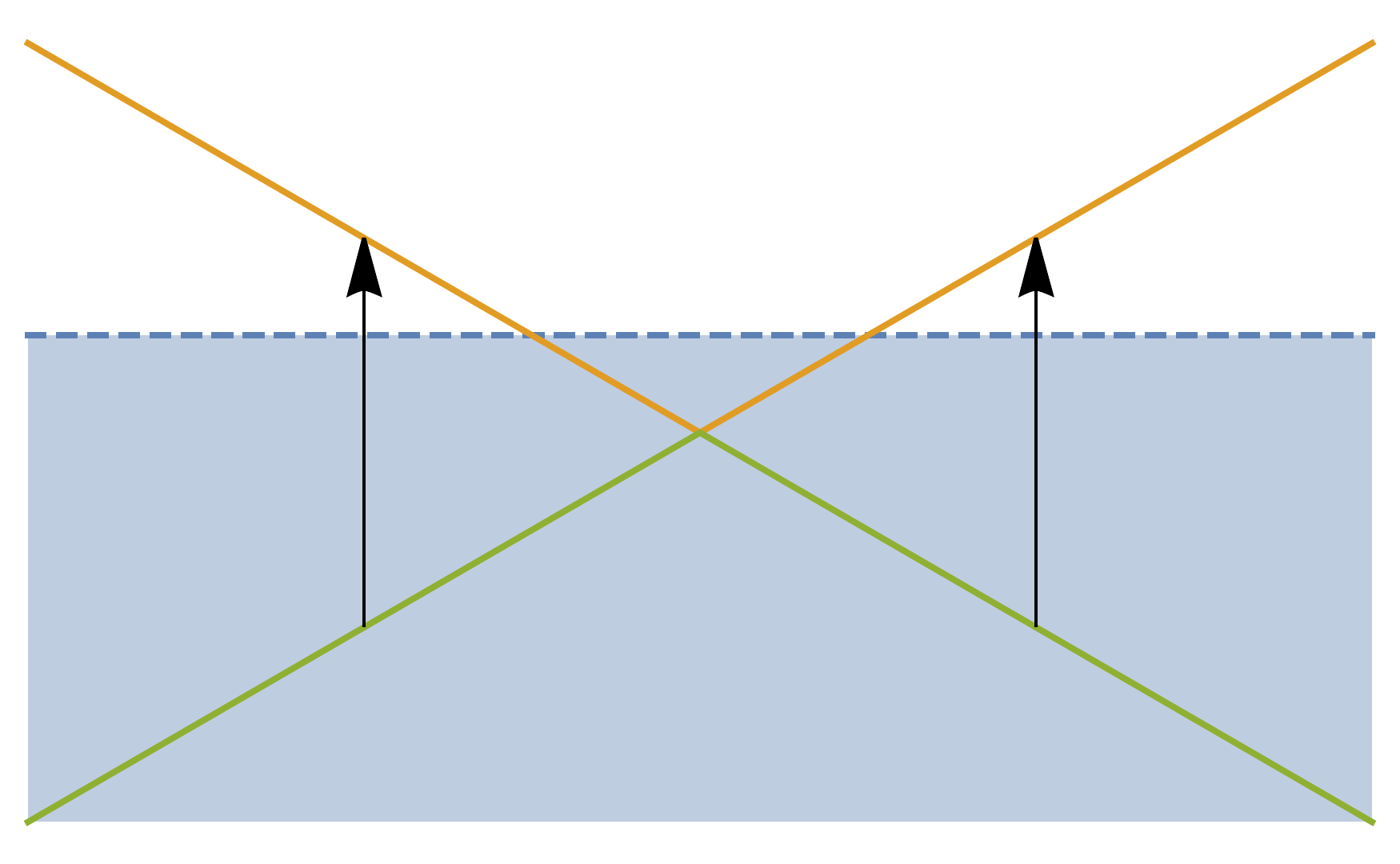}  \includegraphics[scale=0.24]{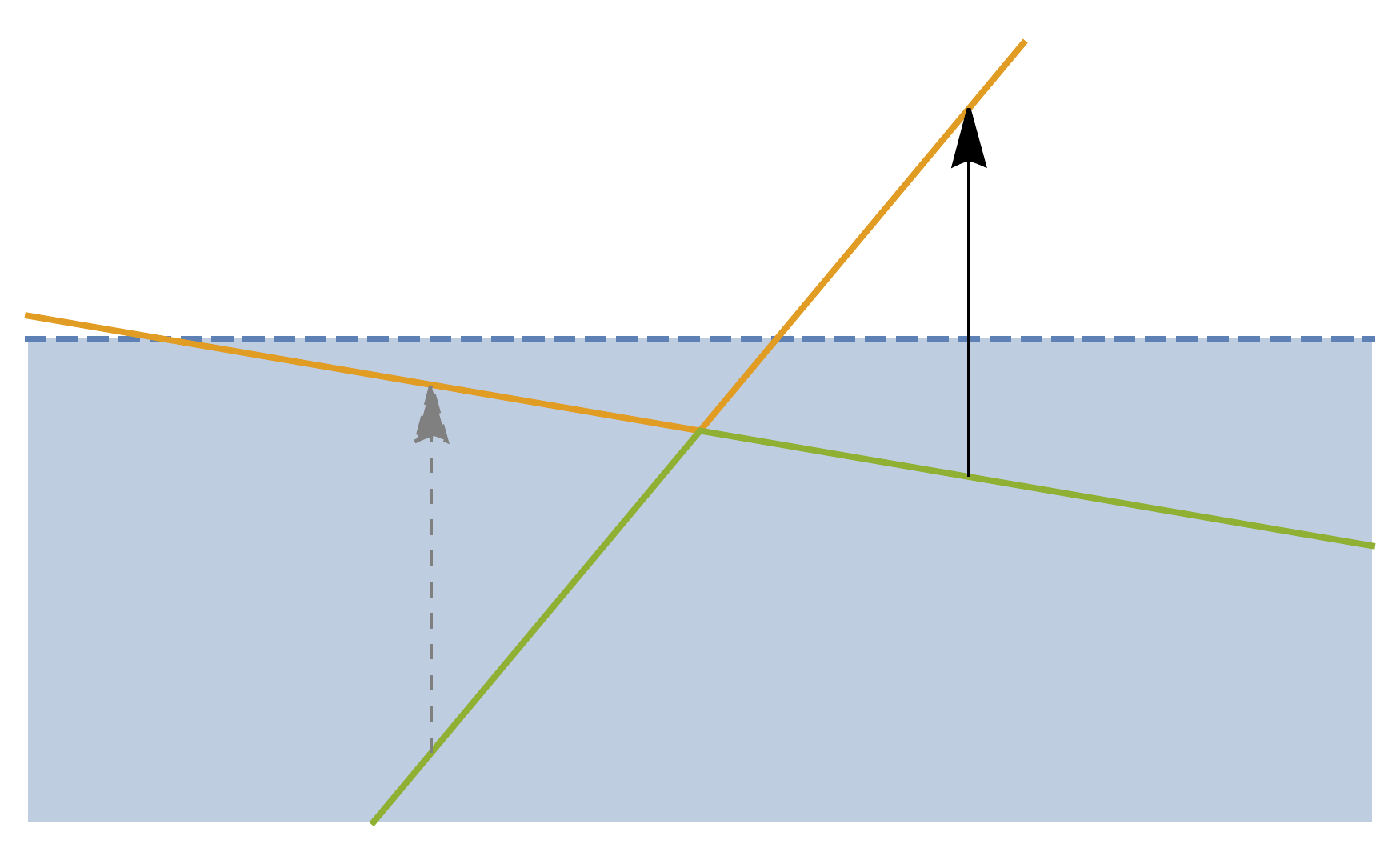}
\caption{Comparison of transitions in an untilted and tilted Weyl cone. In an untilted cone, the initial and final states are related by $CP$ and there is no HME. In the tilted cone, the initial and final states are not related by $CP$, and it is possible to partially blockade the cone. This allows the HME current to be non-zero.}
\label{NormalHME}
\end{figure}

In the HME, linearly polarized light is predicted to produce a current in the presence of a magnetic field \cite{YutaHME}:
\begin{equation}
\vec{j} = \beta_m \epsilon^2 \vec{B}
\end{equation}
where $\vec{\epsilon}$ is the polarization vector. In general, the coefficient $\beta_m$ is a rank-4 tensor. While current is odd under both $P$ and $C$, a magnetic field is even under $P$ and odd under $C$. Thus, the coefficient $\beta_m$ is odd under $P$, even under $C$, and odd under $CP$. 
Consequently, for a Weyl material to exhibit the HME, the (approximate) $CP$ symmetry must be broken by introducing a factor which affects electrons and holes differently, such as a tilt, non-linear terms, or coupling to other bands.
A non-zero Fermi energy is not sufficient to exhibit the CME in an otherwise $CP$-symmetric Weyl material because a transition produces an electron and a hole that are related to each other by $CP$; therefore, this pair cannot contribute to a $CP$-odd coefficient. 
Since a Pauli blockade is uniform for all directions (see Fig.~\ref{NormalHME}) for an untilted Weyl cone, it does not affect the symmetry analysis.

In Ref.~\cite{YutaHME} a Pauli blockade on a tilted Weyl cone was suggested to realize the HME. In a tilted cone, one side has fast electrons and slow holes, and the other has slow electrons and fast holes. A finite Fermi level will blockade one side of the cone, allowing, for example, only transitions that create fast electrons and slow holes, as shown in Fig.~\ref{NormalHME}. This breaks the $CP$ symmetry connecting electrons and holes of opposite chiralities and allows the HME. 

In multifold fermions, because of the lack of tilt, there is still an approximate $CP$ symmetry, which can be seen from Eq~(\ref{eq:HMF}): since $k$ flips sign under $P$ and $S$ remains invariant, the coefficient $\chi v_0$ is odd under $P$. Since holes have opposite energy, momentum, and angular momentum as electrons, $\chi v_0$ also flips sign under $C$, and therefore is invariant under $CP$.
Thus, following the analysis of Weyl fermions, one would naively expect the HME to be absent for chiral multifold fermions with approximate $CP$ symmetry.
However, we will now show by explicit calculation that this is not the case, as long as the Fermi level is not exactly at the band-crossing point.
Heuristically, the non-vanishing HME in the presence of $CP$ symmetry results because there is a unique type of Pauli blockade possible for multifold fermions that is not possible for spin-1/2 Weyl fermions, as shown in Fig.~\ref{blockade}.

\section{HME in Multifold Fermions}\label{secHME}
\label{sec:HMEMF}
\begin{figure}
\centering
\includegraphics[scale=0.24]{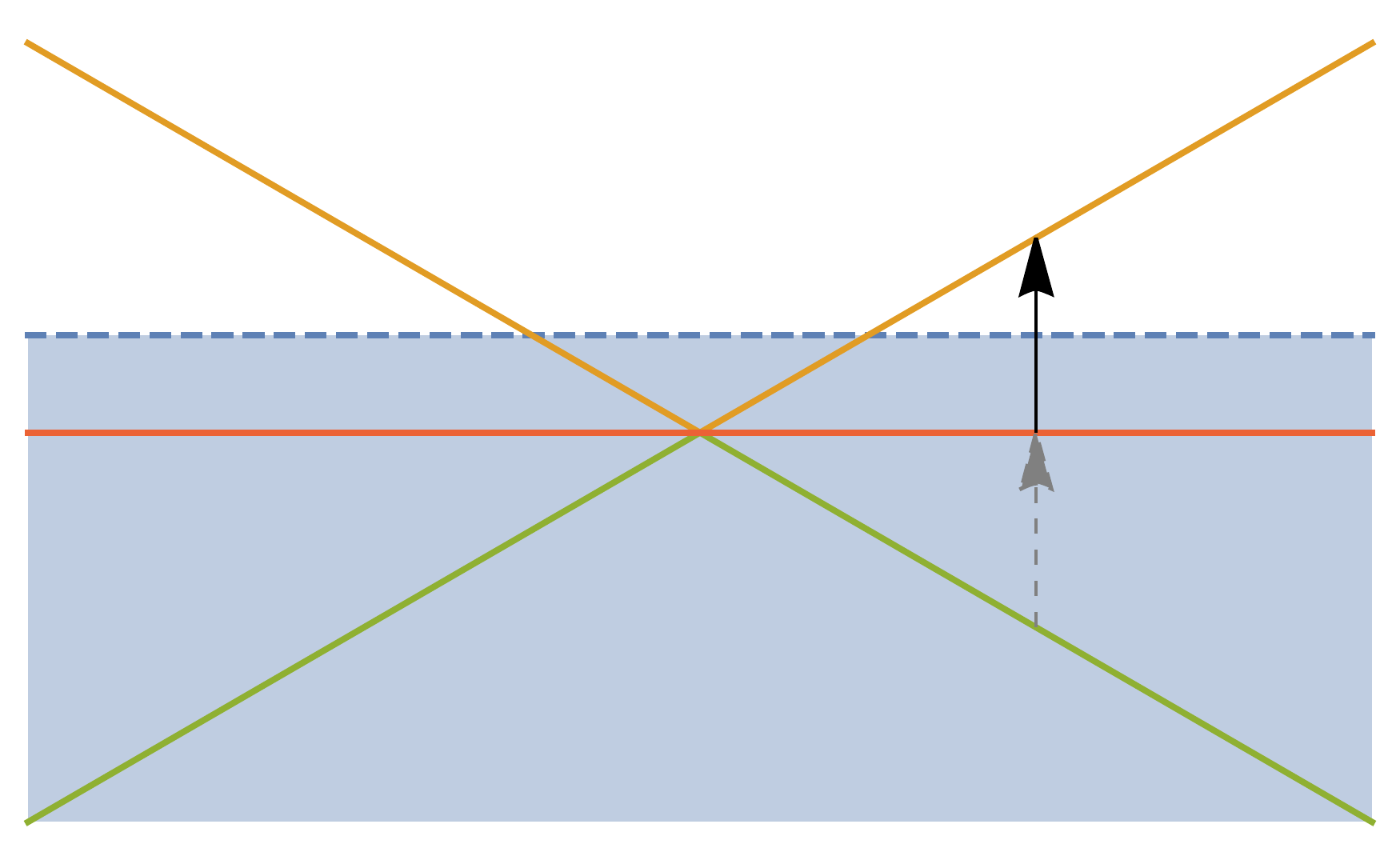}  \includegraphics[scale=0.24]{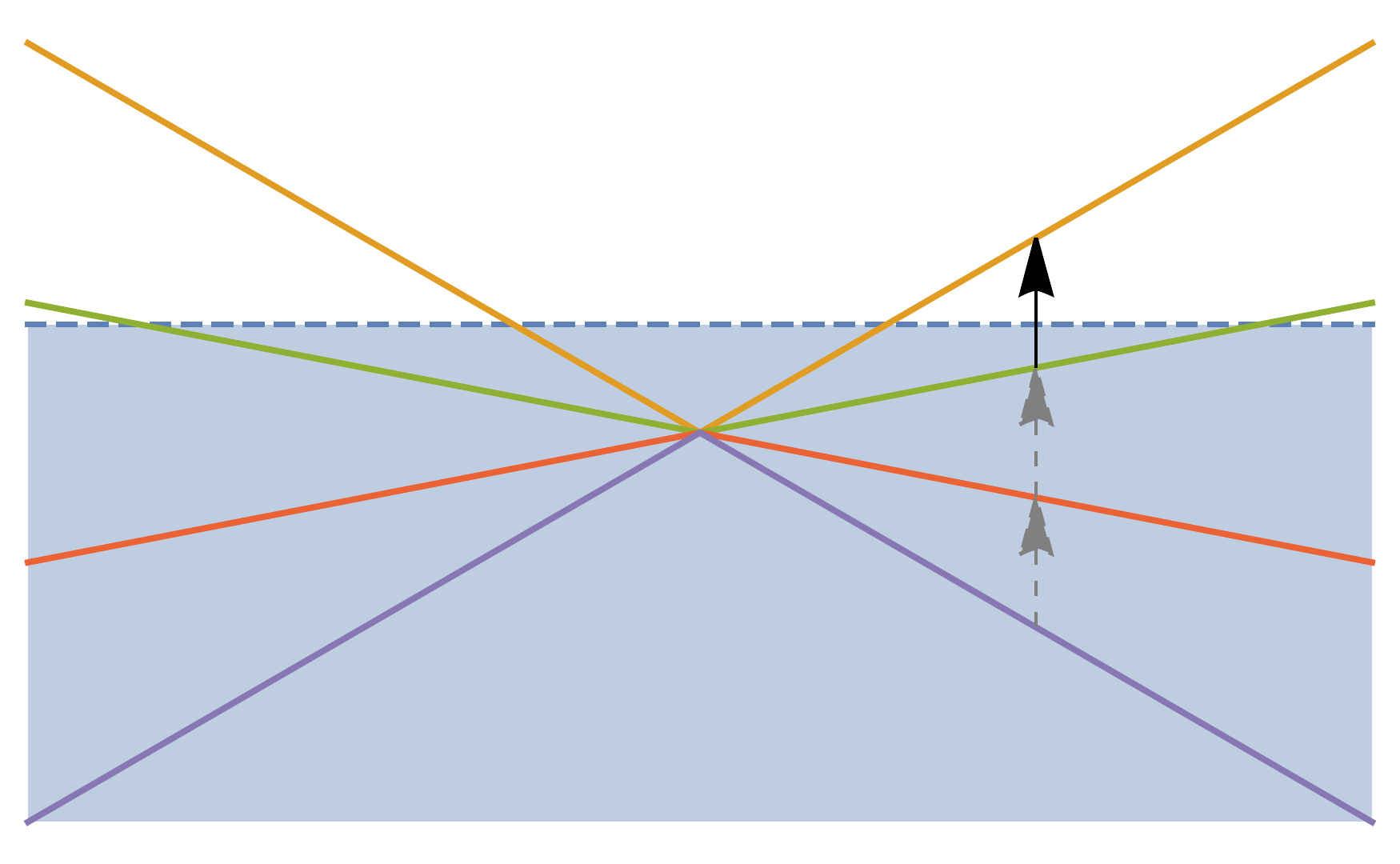}
\caption{Pauli blockade in a spin-1 (left) and spin-3/2 (right) multifold fermion. Solid black arrows indicate allowed transitions from occupied to empty bands, while dashed arrows indicate blockaded transitions. The shaded blue region indicates the Fermi sea. The Pauli blockaded multifold fermions exhibit the HME.}
\label{blockade}
\end{figure}

While multifold fermions described by Eqs~(\ref{eq:HMF}) and (\ref{eq:DHMF}) have $CP$ symmetry, \textit{transitions} between bands with different $|S_k|$ break $CP$ symmetry.
For example, as shown in Fig.~\ref{blockade}, if the Fermi level is such that transitions between bands with $S_k=m$ and $S_k = n \neq \pm m$ are allowed, while transitions to bands with $S_k =-m,-n$ are forbidden, then $CP$ symmetry is ``maximally'' broken compared to tilted simple Weyl cones.
Thus, as long as the Fermi level is not at the charge neutrality point of the multifold fermion, the HME will be present, even though the low-energy theory has $CP$ symmetry. 
This is very different than the situation for Weyl fermions described in the previous section, where the $CP$ symmetry of the Weyl cone must be explicitly broken to exhibit the HME, regardless of the Fermi level.

We now calculate the magnitude of the HME for a multifold fermion.
The effect of chiral Landau levels due to an external magnetic field on the velocity of fermions and the density of phase space can be modeled semiclassically by Chiral Kinetic Theory (CKT) \cite{BerryCorrection,SonCKT,YinCKT,son2013chiral,GolubCKT}, which prescribes:
\begin{align}\label{CKT}
\vec{v} &\to   \frac{\vec{v}+(\vec{v}\cdot \vec{\Omega}) e \vec{B}}{1+e\vec{\Omega}\cdot  \vec{B}}  \nonumber\\
d^3 k &\to (1+e\vec{\Omega}\cdot  \vec{B}) d^3 k
\end{align}
where the unperturbed velocity is $\vec{v} = \nabla_k\ E$, $\vec{\Omega}$ is the Berry curvature and $\vec{B}$ is the applied field.

If the system is in thermal equilibrium, there is no current. The photocurrent occurs because fermions are excited by photons, and have different velocities in their final states compared to initial states. The DC injection current (i.e. current due to optically induced transitions between states of different velocity) in response to a continuous wave is then given by:
\begin{widetext}
\begin{align}\label{general}
\vec{j} = e\sum^{cones}\sum_{i<j}^{bands}\tau  \iint &  \frac{d^3 \vec{k}_i}{(2\pi)^3} (1+e\vec{\Omega}_i\cdot\vec{B})d^3 \vec{k}_j (1+e\vec{\Omega}_j\cdot  \vec{B})\delta(\vec{k}_i-\vec{k}_j)\delta(E_j - E_i - \hbar\omega)\nonumber\\ &\left[\frac{\vec{v}_j + (\vec{v}_j\cdot \vec{\Omega}_j) e \vec{B}}{1+e\vec{\Omega}_j\cdot  \vec{B}} - \frac{\vec{v}_i+  (\vec{v}_i\cdot \vec{\Omega}_i)}{1+e\vec{\Omega}_i\cdot  \vec{B}} e \vec{B}\right] (f_i - f_j) \Gamma_{ij},
\end{align}
\end{widetext}
where $\tau$ is the relaxation time, $f_{i,j}$ is the Fermi distribution function at energy $E_{i,j}$, and $\Gamma_{ij}$ is the transition rate from state $i$ to state $j$, given by Fermi's golden rule:
\begin{equation}
    \Gamma = 2\pi |\langle\psi_i|V_{+\omega}|\psi_j\rangle|^2
    \label{golden}
\end{equation}
where $\psi_{i,j}$ are the initial and final states, $V_{+\omega}$ is the perturbation induced by light.
For an ultrafast pulse shorter than the relaxation time, $\vec{j} = \tau (...)$ in Eq.~(\ref{general}) is replaced by $\frac{d\vec{j}}{dt} = (...)$.

In the absence of a magnetic field, circularly polarized light will result in a nonzero photocurrent because its electric field violates time-reversal; this is exactly the CPGE.
Since the electric field of linearly polarized light is time-reversal symmetric, in a material which is also time-reversal symmetric, this integral vanishes for linearly polarized light in the absence of magnetic field. By definition, the helical magnetic effect is defined as the photocurrent contingent on a magnetic field; thus, we now focus only on terms that depend on $\vec{B}$.

In a spherically symmetric multifold fermion, the energy, velocity, and Berry curvature of the band with $\chi S_k = n$  (Chern number $2n\chi$) is:
\begin{align}
E = nv_0k \nonumber\\
\vec{v} = n v_0 \hat{k}
\label{eq:EVO}\\
\vec{\Omega} = n\chi \frac{\hat{k}}{k^2}\nonumber
\end{align}
Eq.~(\ref{general}) then simplifies. The leading (linear order in $\vec{B}$) term is given by:

\begin{align}\label{special1}
\vec{j} = e\sum_{cones}\sum_{m<n}\tau \chi &\int \frac{d^3 \vec{k}}{(2\pi)^3}  \delta((n-m)v_0 k-\hbar\omega)\\& \left[\frac{n^2 v_0}{k^2} e\vec{B} - \frac{m^2v_0}{k^2} e\vec{B}\right](f_m - f_n)\Gamma_{mn},\nonumber
\end{align}
where $\Gamma$ is the unperturbed transition rate. The transition rate 
from lower to upper states for linearly polarized light with $\vec{A} = \vec{A}_{+\omega}\exp(-i\omega t) + \vec{A}_{-\omega}\exp(i\omega t)$ is
\begin{align}
\Gamma_{mn} =\ & 2\pi|\langle\psi_n|ev_0\vec{A}_{+\omega}\cdot\vec{S}|\psi_m\rangle|^2\nonumber\\ 
=\ & 2\pi e^2 v_0^2 A_{+\omega}^2 \sin^2 \theta |S^x_{nm}|^2
\label{eq:Gammamn}
\end{align}
where $\theta$ is the angle between the electric field and the crystal momentum and
\begin{equation}S^x_{nm} = \frac{1}{2}(\delta_{m,n+1}+\delta_{n,m+1})\sqrt{j(j+1)-mn},
\label{eq:Snm}
\end{equation}
which is obtained by the expressing the elements of the spin matrices in the first line of Eq.~(\ref{eq:Gammamn}) in the basis of $\psi_{m,n}$, i.e., the eigenstates of the Hamiltonian in Eq.~(\ref{eq:HMF}). The selection rules defined by $S^x_{nm}$ in Eq.~(\ref{eq:Snm}) only allow transitions with $n-m = \pm 1$. 


Thus, the photocurrent for each cone is
\begin{equation}\label{special2}
\vec{j} = \chi \frac{e^3I\tau}{6\pi\hbar^2\epsilon_0 c}\frac{2e\vec{B}v_0^2}{\hbar\omega^2} \sum_{m,n} (n^2 - m^2) |S^x_{nm}|^2 (f_m - f_n),
\end{equation}
where $I = 2\epsilon_0 c A^2_{+\omega}\omega^2$ is the intensity of the light, and the Fermi distribution function is $f_m = f(m\omega) = 1/[1+\exp((m\omega - \mu)/T)]$, because the $\delta$-function in Eq.~(\ref{special1}) enforces the transition at $k = \omega/v_0(n-m) = \omega/v_0$.

Since the current is summed over all bands, the HME will be non-zero if there are transitions between bands $m$ to $n$, but not between $-m$ to $-n$.
The factor $(n^2 - m^2)|S^x_{nm}|^2$ is $1/2$ for a spin-$1$ fermion and $3/2$ for a spin-$3/2$ fermion.
However, it is zero for a spin-1/2 Weyl cone because $|n| = |m| = 1/2$.
This explains why the HME vanishes for a symmetric, untilted spin-1/2 Weyl fermion.

Note that the HME does not require $\Gamma$ to have any special form. This means it will be non-zero for any non-zero $\Gamma$, including for linearly polarized or even unpolarized light. For a spherically symmetric linear multifold fermion, the photocurrent is always in the direction of the magnetic field, and is completely independent of the linear polarization of light. In the general case, it can have transverse terms and a polarization dependence.

\begin{figure}
    \centering
    \includegraphics[scale=0.5]{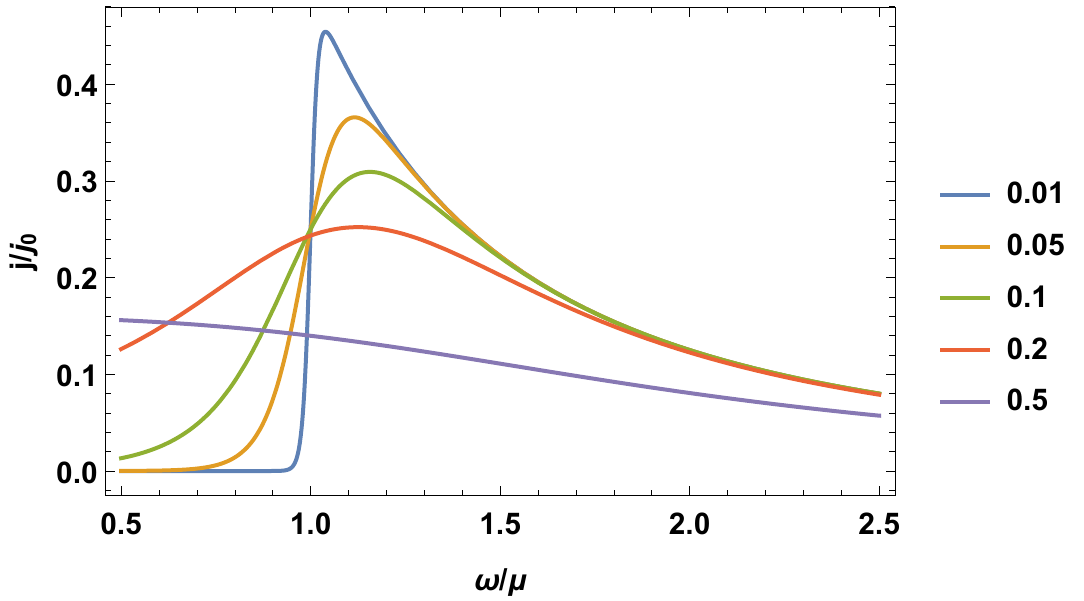}\\ \includegraphics[scale=0.5]{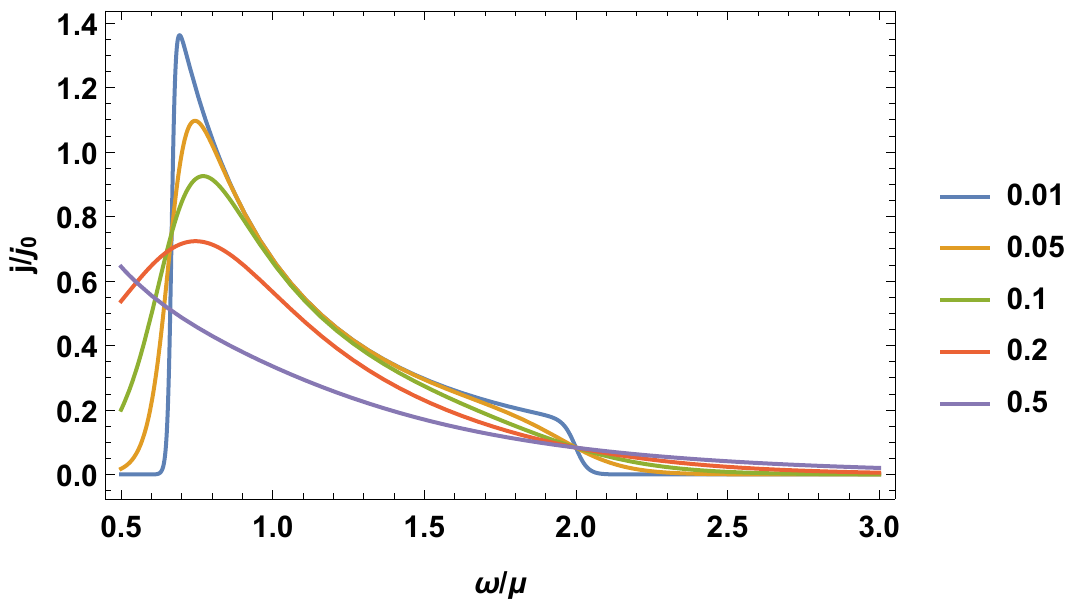}
    \caption{Normalized HME photocurrent vs normalized frequency for a spin-1 cone (top) and a spin-3/2 cone (bottom), for $T/\mu = 0.01, 0.05, 0.1, 0.2$, and $0.5$. The current is normalized in units of $j_0 = \frac{e^3I\tau}{3\pi\hbar^2\epsilon_0 c}\frac{eBv_F^2}{\hbar\mu^2}$ where $v_F$ is the velocity of the fastest band. The frequency is normalized in units of chemical potential $\mu$.
   At $T= 0$, the current is zero until a critical frequency where the upper band is unoccupied at the momentum required for the transition. In the spin-3/2 case, there is also a second critical frequency where the current drops to zero because both upper bands are unoccupied at the momentum required for the transition.}
    \label{linearHME}
\end{figure}

The normalized photocurrent vs normalized frequency are plotted in Figure \ref{linearHME} at different temperatures for both a spin-1 cone and a spin-3/2 cone.
For the spin-1 fermion, the HME is suppressed at low frequencies by a complete Pauli blocade. At frequencies above the chemical potential $\mu$, transitions from the middle band (which has no Berry curvature) to the upper band are allowed, and there is an HME current, which drops as the inverse square of frequency according to Eq.~(\ref{special2}).

For the spin-3/2 fermion, there are fast bands with velocity $3v_0/2$ and slow bands with velocity $v_0/2$. At frequencies less than $2\mu/3$, all transitions are blockaded, and there is no HME. Above this frequency, transitions are allowed because the slower band remains filled, while the faster band is empty.
Above $2\mu$, transitions from the slow to fast band are again suppressed because both bands are empty. 
Transitions from the lower slow band to the upper slow band are allowed at frequencies above $2\mu$, but they do not contribute to the HME as these bands have opposite Berry curvature and hence $n^2 = m^2$.
For frequencies between $2/3\mu$ and $2\mu$, the HME current decreases as the inverse square of frequency according to Eq.~(\ref{special2}).

At finite temperature, the frequency cutoffs described above become smoothed by the Fermi distribution function.

If the Fermi level is below the band degeneracy point, the photocurrent behaves similarly up to a minus sign because the charge carriers would be holes, not electrons.

The quantized CPGE can also be derived by integrating Eq~(\ref{general}) for circularly polarized light (i.e. $A_{+\omega} \sim \hat{x} + i \hat{y}$); the magnitude of the photocurrent is $\frac{e^3I\tau}{6\pi\hbar^2\epsilon_0 c}  |S^x_{nm}|^2$. The multifold HME has a relative factor of $\frac{2eBv^2}{\hbar\omega^2} (n^2 - m^2)$ vs the quantized CPGE. This means it is inversely proportional to the number of Landau levels involved in the photoexcited transitions.

While the quantized CPGE does not require a magnetic field and depends on the direction of the incident light and its \textit{circular} polarization, the multifold HME is roughly independent of the direction and polarization of the light; it is (approximately) parallel or antiparallel to the external magnetic field.

\section{Non-linearity and Transitions from Multifold Fermions to Other Bands}\label{secNonLin}

In a crystal, multifold fermions do not have full spherical symmetry, only the symmetry of the little group at their crystal momentum.
Non-linear terms will generically be present and break spherical symmetry. Even in cubic crystals, spin-1, spin-3/2, and double spin-1 fermions without time-reversal symmetry can also have \textit{linear} terms that break the approximate spherical symmetry\cite{CanoMultifold,FlickerMultifold}, such as:
\begin{equation}
    a (k_x S_x^3 + k_y S_y^3 + k_z  S_z^3)
\end{equation}
for spin-3/2 fermions;
\begin{equation}
    a \tau_2 (k_x \{S_y,S_z\} + k_y \{S_z,S_x\} + k_z  \{S_x,S_y\})
\end{equation}
for double spin-1 fermions; and
\begin{equation}
     a (k_x \{S_y,S_z\} + k_y \{S_z,S_x\} + k_z  \{S_x,S_y\})
\end{equation}
for spin-1 fermions not at TRIMs or in crystals with broken time-reversal.

In addition, transitions to other bands outside the multifold fermion are possible.
All these effects will contribute to the HME photocurrent, 
causing it to deviate from the idealized form in Eq.~(\ref{special2}).
In these cases, the HME photocurrent can be calculated from the general formula in Eq.~(\ref{general}). 

\begin{figure}
    \centering
    \includegraphics[scale=0.25]{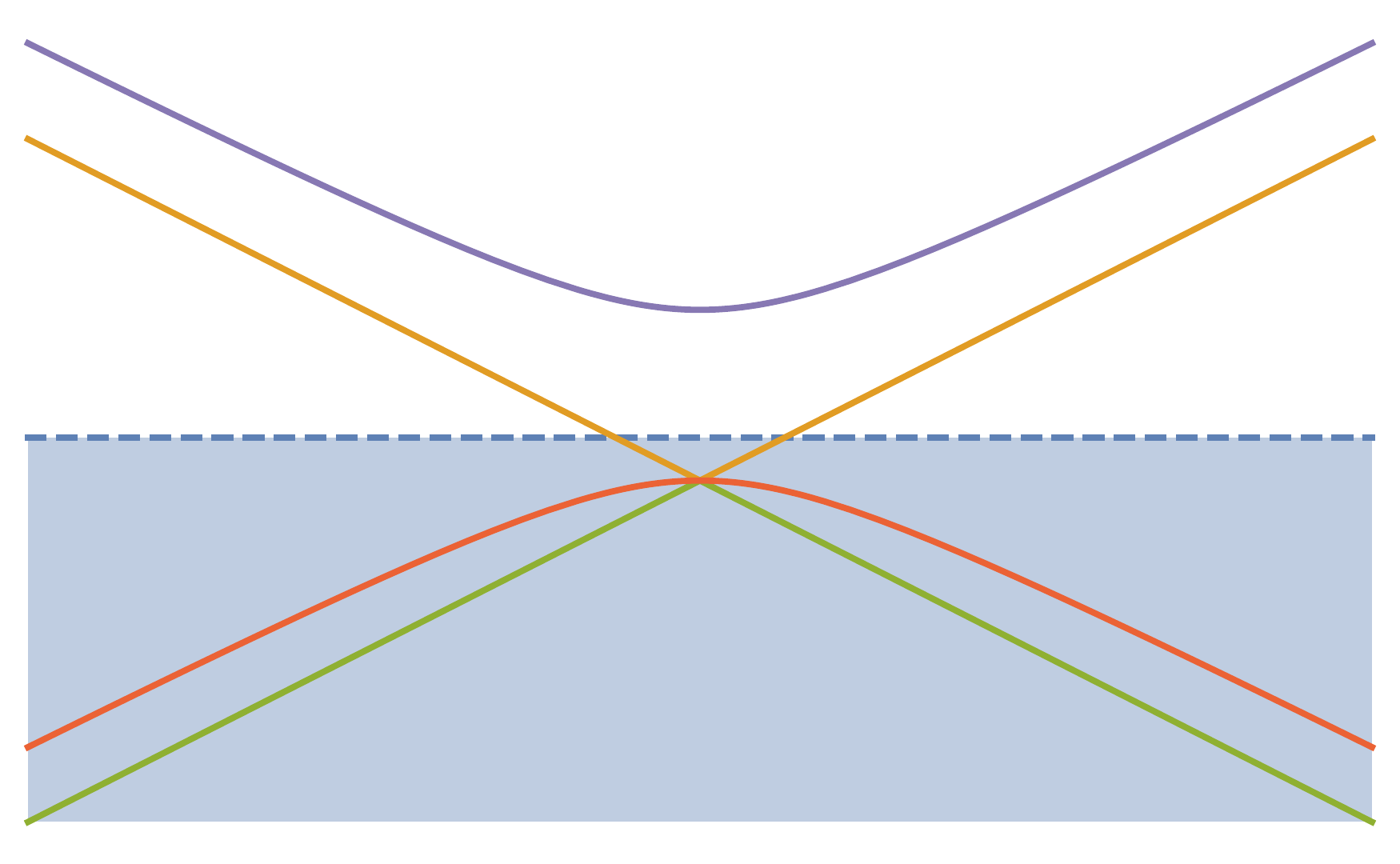}\\
    \includegraphics[scale=0.5]{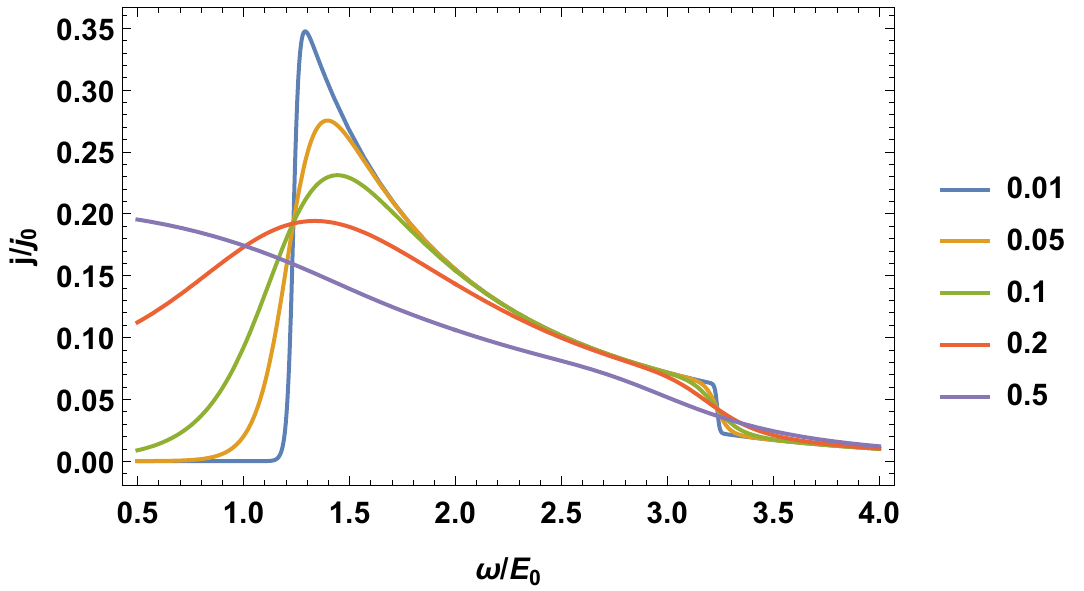}
    \caption{The dispersion relation (top) and normalized photocurrent vs normalized frequency (bottom) for the double spin-1/2 fermion split into a spin-1 fermion and trivial band, described by the Hamiltonian in Eq.~(\ref{bandgap}). 
    The photocurrent is normalized in units of $\frac{e^3I\tau}{3\pi\hbar^2\epsilon_0 c}\frac{eBv_F^2}{\hbar E_0^2}$ while the frequency is normalized in units of $E_0 = \Delta/4$. The chemical potential is taken to be $\mu = 0$. The polarization is along the magnetic field. The normalized temperature is $T/E_0 =  0.01, 0.05, 0.1, 0.2$, and $0.5$. At $T=0$, the current is zero for small frequencies. Above a critical frequency, the Pauli blockade is lifted only for transitions from trivial to chiral bands, and there is a current. Above a second critical frequency, the Pauli blockade is also lifted for transitions from chiral to trivial bands, and the current is \textit{approximately} cancelled, but is not exactly zero.}
    \label{splitCME}
\end{figure}

To illustrate the effects of non-linearity and transitions to trivial bands, we consider a double spin-1/2 fermion described by the Hamiltonian in Eq~(\ref{eq:DHMF}) plus a small perturbation of the form $\sum\tau_i\sigma_i$ that splits it into a spin-1 and a trivial fermion, as shown in Figure~\ref{splitCME}. Thus, this model is an example of a spin-1 fermion with both non-linear terms and transitions to another band.
This system can be described by a Hamiltonian of the form
\begin{align}\label{bandgap}
H =  v_0 i &\begin{pmatrix}0      & k_x  & k_y  & k_z  \\
-k_x & 0      & k_z  & -k_y \\
-k_y & -k_z & 0      & k_x  \\
-k_z & k_y  & -k_x & 0 \end{pmatrix}\\
+  \frac{\Delta}{4} &\begin{pmatrix}3 & 0  & 0  & 0  \\
0 & -1 & 0  & 0  \\
0 & 0  & -1 & 0  \\
0 & 0  & 0  & -1\end{pmatrix}\nonumber
\end{align}

We plot the dispersion relation and HME current vs frequency for this system at different temperatures in Fig.~\ref{splitCME}. We consider the Fermi level to be between the Weyl node and the upper trivial band, i.e. between $-E_0$ and $3E_0$ where $E_0 = \Delta/4$. At low frequencies, all transitions are blockaded, and there is no photocurrent. Above $\omega = \mu  -E_0 + \sqrt{4E_0^2 + (\mu+E_0)^2}$, transitions from the lower trivial band to the upper chiral band are not blockaded, while those from the lower chiral band to the lower trivial band are blockaded, which marks the onset of the photocurrent. The system is similar to an isolated and symmetric spin-1 fermion (Fig.~\ref{linearHME}) and there is a large photocurrent. Above $\omega = \mu  +E_0 + \sqrt{4E_0^2 + (\mu+E_0)^2}$, transitions from the chiral bands to the upper trvial band are also allowed, which contribute a term with opposite sign to the photocurrent, which corresponds to the sharp drop in photocurrent. However, the cancellation is not exact because of non-linearity. At higher frequencies, the photocurrent is strongly suppressed because the system resembles a untilted double spin-1/2 cone. The photocurrent also has a very weak dependence on the polarization of light due to the non-linearity.

\section{Magnetic Photocurrents from Landau Levels}\label{secLandau}

\begin{figure}
    \centering
    \includegraphics[scale=0.075]{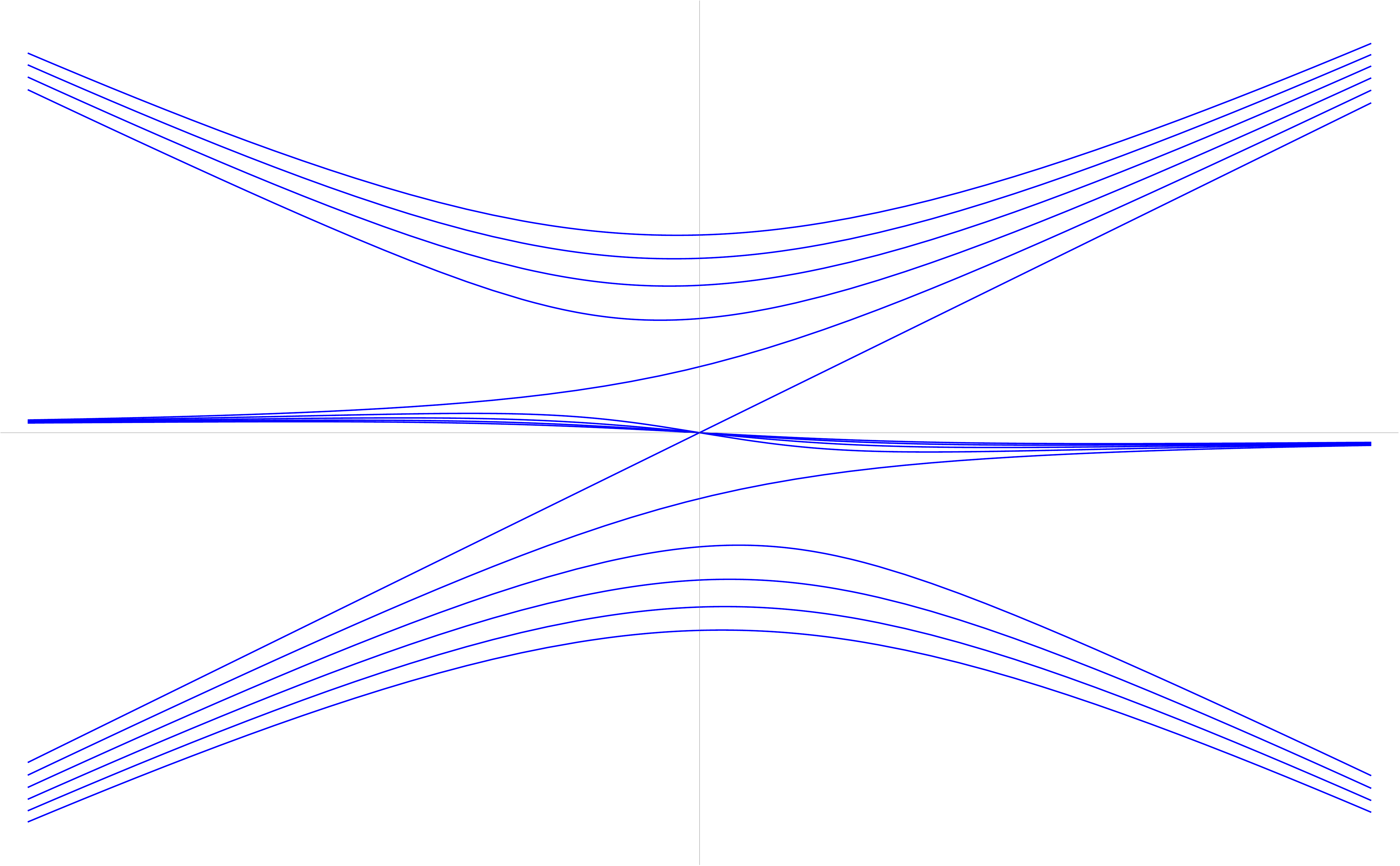}
    \includegraphics[scale=0.0675]{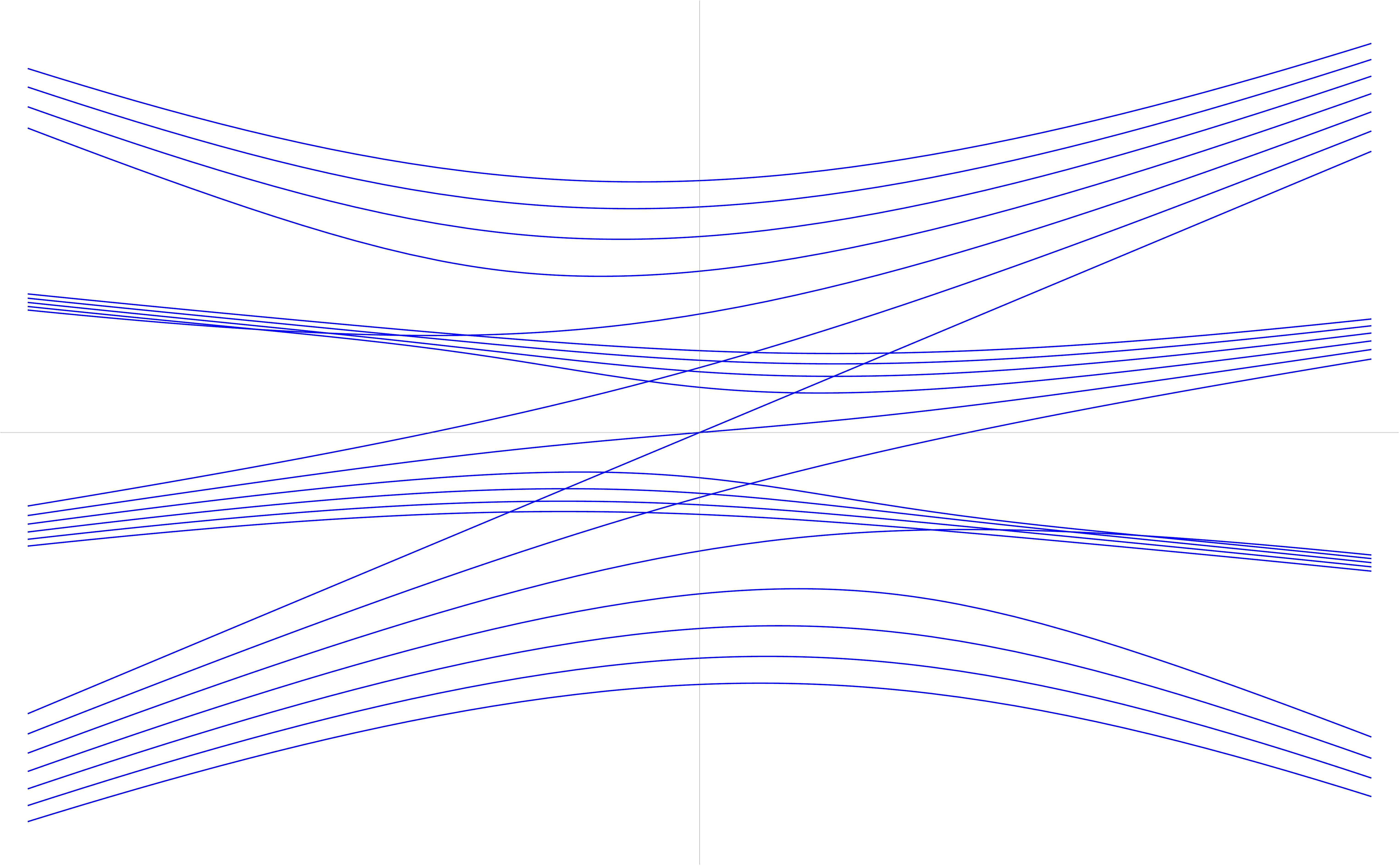}
    \caption{The chiral Landau levels and first five achiral Landau levels for a spin-1 fermion (left) and a spin-3/2 fermion (right).}
    \label{LL}
\end{figure}

The factor of $(n^2 - m^2) \frac{2eBv_0^2}{\omega^2}$ that appears in Eq.~(\ref{special2}) can also be understood from the Landau level spectrum of a multifold fermion.
In a magnetic field, chiral fermions exhibit chiral Landau levels, which propagate only in one direction.
Each band contributes a number of chiral Landau levels equal to its Chern number.
The Landau levels of a spin-1 fermion and a spin-3/2 fermion are illustrated in Fig.~\ref{LL}. 
The chiral anomaly can be interpreted as a consequence of these unpaired chiral Landau levels.

The effects of the unpaired Landau levels in the semiclassical calculation are captured by the deformation of phase space in Eq.~(\ref{CKT}). When the temperature or inverse scattering time is larger than the Landau splitting, we can ignore the quantum oscillations and focus only on the deformation of phase space \cite{GustavoOscillations}.
Here we show that the same scaling can be obtained for a spherically symmetric multifold fermion by counting the density of states in each chiral Landau level.

When the fermions are in thermal equilibrium, the total current vanishes because the Chern numbers of left and right handed cones cancel. 
However, a net current is possible when we disturb the distribution function by shining light.

Consider a spherically symmetric linear multifold fermion in a magnetic field along the $\hat{z}$ direction.
A band with $S_k = n$ contributes $2n$ chiral modes, corresponding to its Chern number.
In the semiclassical limit, the velocity of each chiral mode is $n\chi v_0$ along the direction of the magnetic field, the same as the speed of the unperturbed band (see Eq.~(\ref{eq:EVO})). 
The density of states per area within each Landau level in the $x-y$ plane is $eB/2\pi$; each Landau level corresponds to an area of $2\pi eB$ in the $k_x-k_y$ plane. 
The fermions that participate in transitions excited by light of frequency $\omega$ are located on a sphere in momentum space of radius $k = \omega/v$.
The number of Landau levels involved in these transitions is proportional to the cross section $\pi k^2$ of this sphere, so that $n_{LL} = \pi k^2/2\pi eB = \omega^2/2eBv_0^2$. 
The total number of fermions participating in transitions is proportional to the surface area of the sphere $4\pi k^2$, while the number of fermions in each chiral Landau level on the sphere (which have $k_x, k_y \sim 0$, because at fixed $k^2$ they have maximal $|k_z|$ for their band, and therefore minimal $k_x, k_y$) is proportional to the area of the Landau level projected onto the sphere, which for $\vec{k}\sim k\hat{z}$ is the same as the area of the Landau level $2\pi eB$. 
The fraction of fermions belonging to each chiral Landau level is $2\pi eB/4\pi k^2 = eB/2k^2 = eBv_0^2/2\omega^2$.
Because of these unpaired chiral modes, the average velocity along the magnetic field of the fermions in that band participating in the transition is 
\begin{equation}
   \langle \vec{v}_n \rangle = 2n\times n\chi v_0 \times \frac{e\vec{B}v_0^2}{2\omega^2} = n^2\chi \frac{e\vec{B}v_0^2}{\omega^2} v_0.
\end{equation}
 If fermions are excited from a band with $S_k = m$ to one with $S_k = n$, the average change in velocity is 
 \begin{equation}
\langle\Delta\vec{v}\rangle = \chi (n^2 - m^2) \frac{e\vec{B}v_0^2}{\omega^2} v_0
 \end{equation}
 along the magnetic field, which explains the scaling with $n^2-m^2$ and $B$ in Eq.~(\ref{special2}).

We now compare the HME and the CPGE. In the CPGE, the magnitude of the change in velocity for each transition is $v_0$. The rate of transition for circularly polarized light, from Eq.~(\ref{golden}), is proportional to $(1+\cos\theta)^2$ where $\theta$ is the angle between the momentum of the fermion and the angular momentum of light; the change in velocity projected along the angular momentum of light is $v_0 \cos\theta$.
Therefore, the change in \textit{velocity}, averaged over the whole sphere is
\begin{equation}
    \langle\Delta\vec{v}\rangle  = \frac{\int (1+\cos\theta)^2 \cos\theta \ 2\pi \sin\theta d\theta}{\int (1+\cos\theta)^2  \ 2\pi \sin\theta d\theta}v_0 = \frac{1}{2}v_0
\end{equation}
Therefore, the ratio of average change in velocity per excitation in the HME vs the CPGE is $(n^2 - m^2) \frac{2eBv_0^2}{\omega^2}$, which is precisely the ratio between the HME and the CPGE in Eq.~(\ref{special2}).
The factor $\frac{2eBv_0^2}{\omega^2} = n_{LL}^{-1}$ is the inverse number of Landau levels involved in the transitions.
This scaling obtained by counting the states in each chiral Landau level agrees with the calculation using chiral kinetic theory in Sec.~\ref{secHME}.

\section{Material realizations: Cobalt Silicide and Rhodium Silicide}\label{secMats}

The HME photocurrent derived in Sec.~\ref{secHME} should be present in the known multifold fermion materials in the B20 family, such as RhSi, CoSi, and AlPt \cite{tang2017multiple,HasanRhSi,sanchez2019topological,schroter2019chiral,RhSiCurrent,xu2020optical}.
These materials are in space group $P2_13$ (SG 198) with chiral tetrahedral symmetry. 
Ignoring spin-orbit coupling (SOC),
they exhibit a double spin-1 fermion, separated from a double trivial fermion by a large energy, at $\Gamma$ and a quadruple spin-1/2 fermion at $R$.
Thus, the multifold fermions are maximally separated in momentum space and not related to each other by even an approximate symmetry. 
The HME photocurrent in this case will come predominantly from the double spin-1 fermion at $\Gamma$, since the spin-1/2 fermions at $R$ do not contribute to the HME because symmetry forbids them from having a tilt.
For small frequencies, the HME photocurrent from $\Gamma$ will be approximately twice the contribution from a single spin-1 fermion, derived in Sec.~\ref{sec:HMEMF}. At larger frequencies, the quadratic dispersion of the middle band (which is flat to linear order, as shown in Fig.~\ref{blockade}) will cause the photocurrent to deviate from its idealized value; nonetheless, it should follow the general trend of the photocurrent plotted in Fig.~\ref{linearHME} (upper), where the photocurrent is nearly zero until a finite onset frequency (necessary to overcome the Pauli blockade) and then decreases.

SOC splits the bands at $\Gamma$ into a spin-3/2 fermion and a Weyl fermion and splits the bands at $R$ into a double spin-1 fermion and two quadratic bands, essentially two copies of the model described in Eq.~(\ref{bandgap}). 
In general, SOC will cause the HME photocurrent to deviate from its idealized form in Eq.~(\ref{special2}), and the general formula in Eq.~(\ref{general}) must be applied.
For mid-infrared light, the frequency is larger than the spin-orbit coupling, but significantly smaller than the separation at $\Gamma$, and we can approximate the HME current by considering only the double spin-1 fermion at $\Gamma$, and ignoring terms that break spherical symmetry.

We now compare the magnitude of the HME photocurrent to that of the CPGE, which has already been observed in RhSi \cite{RhSiCurrent,ni2020linear} and CoSi \cite{CoSiCurrent}. 
As discussed above, the HME photocurrent will come predominantly from the double spin-1 fermion at $\Gamma$.
In a material with a double spin-$1$ cone with Fermi velocity $3\times 10^5\  \mathrm{m/s}$, with the lower and middle bands fully occupied, but the upper band fully unoccupied, excited by light of energy $100\ \mathrm{meV}$, in a magnetic field of $5\ \mathrm{T}$, along the surface of the crystal, there would be $\sim 16$ Landau levels at the excitation energy. Ignoring the effect of the angle of incidence, the photocurrent would be $\sim 0.06$ times the CPGE contribution of the same cone. 
However, depending on the energy of light and the chemical potential, the CPGE could cancel, because the contributions to the CPGE from $\Gamma$ and $R$ enter with opposite sign, while the HME photocurrent does not have this putative cancellation because the spin-1/2 fermions at $R$ do not contribute.

Returning to the importance of the incident angle, the CPGE is always parallel (or anti-parallel) to the angular momentum of light, so the \textit{observed} CPGE, i.e. the component along the surface of crystal, has a factor of the sine of the angle of refraction. If the angle of refraction is $10^\circ$, the CPGE will have a factor of $\sim 0.16$, and the HME current will be $\sim 0.4$ times the CPGE current. Note that the HME current would always be in the direction of the magnetic field and roughly independent of the linear polarization (it is perfectly independent if the bands are perfectly linear).

\section{Discussion}\label{secDisc}

We have demonstrated that the Helical Magnetic Effect can occur in multifold Weyl fermions, as long as the Fermi level is not exactly at the degeneracy point. Unlike simple Weyl fermions, in multifold fermions this effect occurs even in the idealized limit, i.e., it does not require tilt, deviations from linearity or breaking of spherical symmetry.
We derived the HME photocurrent for an ideal multifold fermion in Eq.~(\ref{special2}), which is plotted in Fig.~\ref{linearHME} for a spin-1 and spin-3/2 fermion.
To demonstrate the effect of perturbations beyond the ideal case, we also considered a double spin-1/2 fermion split into a spin-1 fermion and a trivial band; the resulting HME photocurrent in this system is shown in Fig.~\ref{splitCME}.

The HME could be distinguished from other contributions to the photocurrent, such as the CPGE by its dependence on polarization and magnetic field. The HME would be linear in magnetic field, and roughly independent of the linear or circular polarization of incident light.

For the same intensity of incident light, the HME is of the order of the CPGE divided by the number of Landau levels at the energy of the incident light. However, unlike the CPGE, the HME will be observable even at normal incidence, and for linearly polarized light, and not be suppressed by the sine of the refracted angle.

We predict that the HME is observable in materials known to exhibit multifold fermions such as RhSi, CoSi, and AlPt.
We estimate its magnitude will be within an order of magnitude of the CPGE photocurrent depending on the chemical potential, magnetic field, and incident angle of light.

\begin{acknowledgments}
We thank Barry Bradlyn, Dmitri Kharzeev, Nitish Mathur, Evan Philip, and Liang Wu for useful and stimulating discussions. 
This work was supported in part by the U. S. Department of Energy
under Awards DE-SC-0017662 (S. K.) and by the National Science Foundation under award DMR-1942447 (J. C.).
J.C. also acknowledges the support of the Flatiron Institute, a division of the Simons Foundation.
J. C. and S. K. acknowledge the support of an OVPR Seed Grant from Stony Brook University.
\end{acknowledgments}

\bibliographystyle{apsrev4-1}
\bibliography{refs}

\end{document}